# Pervasive Image Computation: A Mobile Phone Application for getting Information of the Images

Reza Rahimi, J Hengmeechai
Faculty of Engineering,
University of Regina,
Regina, SK, Canada.

*Abstract*—Although many of the information processing systems are text-based, much of the information in the real life is generally multimedia objects, so there is a need to define and standardize the frame works for multimedia-based information processing systems. In this paper we consider the application of such a system namely pervasive image computation system, in which the user uses the cellphone for taking the picture of the objects, and he wants to get some information about them. We have implemented two architectures, the first one, called *online* architecture, which the user sends the picture to the server and server sends the picture information directly back to him. In the second one, which is called *offline* architecture, the user uploads the image in one public image database such as Flickr and sends the ID of the image in this database to the server. The server processes the image and adds the information of the image in the database, and finally the user can connect to the database and download the image information. The implementation results show that these architectures are very flexible and could be easily extended to be used in more complicated pervasive multimedia systems.

## I. INTRODUCTION

We are in the beginning of the new era of the computer and communication technology usually called *Ubiquitous* or *Pervasive* computing and communication. It is considered as the new wave after Internet. The essence of it is the creation of environments saturated with computing and communication, yet gracefully integrated with human users [1]. This era will change and define the new way of life for human being in different layers, for example in health system, banking, friendship, etc. Many pilot works have been done in the academic area and industry for making framework and new applications, such the improvement of educational environment, or health system [1], [2], [3].

To have the suitable platform for pervasive computation, new technologies both in software and hardware level are required and should be designed. The first step toward the pervasive computation is the *distributed* and *mobile* computing [6] [5], which is started from the 1970 and still continues. In this context the computation and communication are distributed among computers and mobiles that makes the fertile platform for design and implementation of new applications. The next wave which is started recently is switching from the web computing concept to the *Semantic Web* [4], in which the semantics of information and services on the web is defined, making it possible for the web to understand and satisfy the requests of people and machines to use the web content pervasively.

In this paper we consider a general application namely pervasive image computation, in which the mobile user uses his cellphone for getting image information and wants to get some information about it. This application is very common in everyday life such that when we take a picture of one flower our purpose is to know the name of flower and the nearest shop to buy. The proposed architecture in this paper could be easily extended to the broader class of multimedia objects. The rest of the paper is organized as follow. In section II the general use case, requirements and proposed architecture for the pervasive image computation is described. The conclusion is made in section III.

## II. PERVASIVE IMAGE COMPUTATION

The multimedia information processing systems have many real life applications. It happens many times that we are listening to the very beautiful music and want to know the name of the composer or the music, or as a student we want to take the picture of the blackboard and find a way to convert the image to the lecture notes. So with having suitable framework, many applications can be implemented. This paper focuses on the pervasive image computation systems and its potential use cases as shown in Fig.1. As it can be seen in use case diagram of the system, many applications could be considered for the system. For example it happens many times that we want to know the estimated size of an object, or in the restaurant we want to know the estimated calorie of the food.

In this paper we have implemented one of these use cases, in which we take a picture of the book barcode and want to know the price and maybe the cheapest one to buy. For the implementation of the system the J2EE, J2ME platforms have been used [8], [9] and the image computation server has been implemented as the web service, using both SOAP and REST [7] architectures. In the next we are going to describe two different architectures of the system.

*A. Online Architecture*

In this scenario the user takes a picture of the barcode and sends it to the server using SOAP or REST protocol. The server processes the image and gets the barcode number. It uses SOAP to use the Amazon web service to get the book information, and sends the information back to the mobile user.

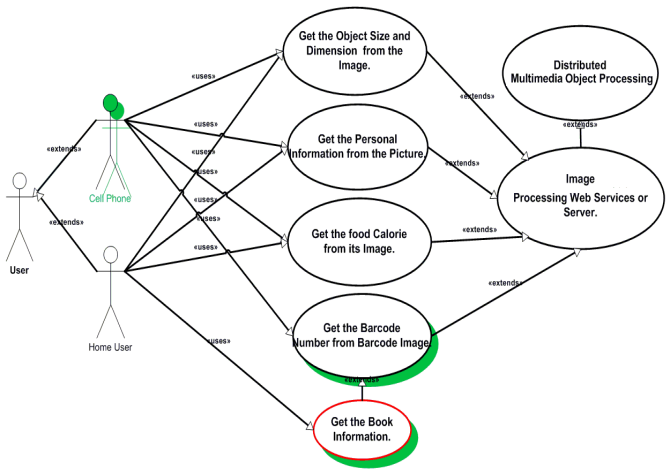

Fig. 1. The use case diagram of the Pervasive Image Computation System.

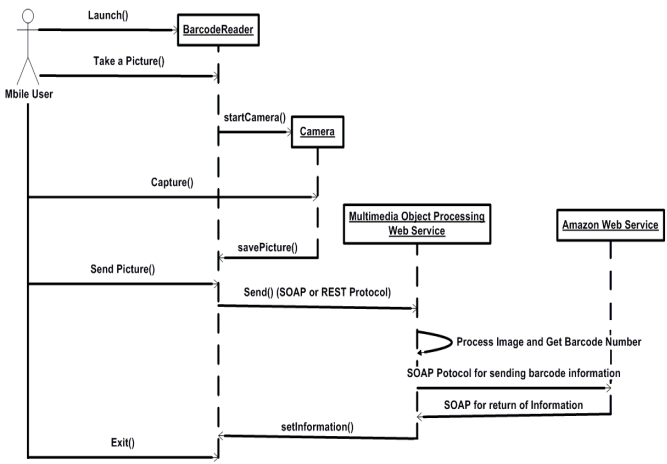

Fig. 2. Online architecture for the sample application when the user wants to get the book information from its barcode.

Fig. 2 shows the sequence diagram of the this case, called online scenario.

### B. Offline Architecture

This architecture is very useful for the system with limited resources. In this case the image database is used as the useful way for storing images and save the communication resources. For this scenario, the mobile user takes a picture of the barcode image and uploads it to Flickr using REST protocol [10] and gets the photo ID. Then the photo ID is sent to the image processing server. The server uses this ID for getting information from Flickr and processes the image to get the barcode. The server uses the barcode for getting the information from Amazon as described in the online scenario. Finally the information is saved as the photo tag in the Flickr. When everything gets ready the server uses SMS message to inform the user that the data is ready in Flickr to pick up. This scenario is called offline and Fig. 3 shows its sequence diagram.

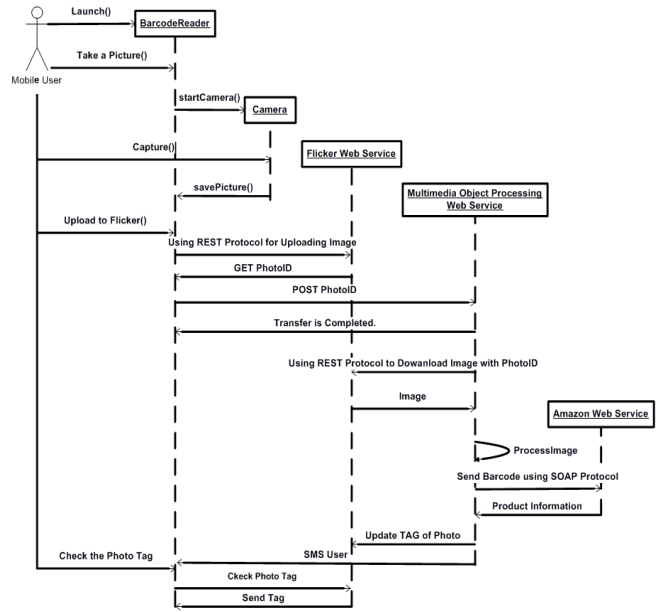

Fig. 3. Offline architecture for the sample application when the user wants to get the book information from its barcode.

### III. CONCLUSION

In this paper the sample application for pervasive image computation system was described and two different architectures were proposed. These architectures could be easily extended to the more general cases in pervasive multimedia computation. We only consider the use case when we need to get information from barcode image. We are now trying to extend the application for estimating the calorie of the food, which could be very helpful in health system for controlling obesity and the other application in which converting the image of handwriting to the text.